\documentclass[prb,twocolumn,showpacs,preprintnumbers,superscriptaddress,amsmath,amssymb]{revtex4}
\usepackage{graphicx}
\usepackage{amsmath}
\usepackage{subfigure}
\usepackage{braket}
\usepackage{color}
\usepackage[dvipsnames]{xcolor}
\usepackage{epstopdf}
\usepackage[normalem]{ulem}

\definecolor{ao(english)}{rgb}{0.0, 0.42, 0.24}

\begin{document}

\title{X-ray absorption of liquid water by advanced \emph{ab initio} methods}

\author{Zhaoru Sun}

\author{Mohan Chen}
\affiliation{Department of Physics, Temple University, Philadelphia, PA 19122, USA}
\author{Lixin Zheng}
\affiliation{Department of Physics, Temple University, Philadelphia, PA 19122, USA}
\author{Jianping Wang}
\affiliation{Department of Physics, Temple University, Philadelphia, PA 19122, USA}
\author{Biswajit Santra}
\affiliation{Department of Chemistry, Princeton University, Princeton, NJ 08544,USA}
\author{Huaze Shen}
\affiliation{International Center for Quantum Materials, School of Physics, Peking University,
 Beijing 100871, China}
\author{Limei Xu}
\affiliation{International Center for Quantum Materials, School of Physics, Peking University,
 Beijing 100871, China}
\author{Wei Kang}
\affiliation{College of Engineering, Peking University, Beijing 100871, China}
\author{Michael L. Klein}
\affiliation{Department of Physics, Temple University, Philadelphia, PA 19122, USA}
\affiliation{Department of Chemistry, Temple University, Philadelphia, PA 19122, USA}
\affiliation{Institute for Computational Molecular Science, Temple University, Philadelphia, PA 19122, USA}
\author{Xifan Wu}
\thanks{$^\ast$Corresponding author. Email:xifanwu@temple.edu}
\affiliation{Department of Physics, Temple University, Philadelphia, PA 19122, USA}
\affiliation{Institute for Computational Molecular Science, Temple University, Philadelphia, PA 19122, USA}

\date{\today}

\begin{abstract}
Oxygen $K$-edge X-ray absorption spectra of liquid water are computed
based on the configurations from advanced \emph{ab initio} molecular dynamics simulations,
as well as an electron excitation theory from the GW method.
One one hand, the molecular structures of liquid water are accurately predicted by including
both van der Waals interactions and hybrid functional (PBE0).
On the other hand, the dynamic screening effects on electron excitation are approximately described by the recently developed
enhanced static Coulomb hole and screened exchange approximation by Kang and Hybertsen [Phys. Rev. B 82, 195108 (2010)].
The resulting spectra of liquid water are in better quantitative agreement with the experimental spectra
due to the softened hydrogen bonds and the slightly broadened spectra originating from the better screening model.
\end{abstract}

%\keywords{}
\maketitle

\section{Introduction}
Water is arguably one of the most important materials on Earth and needs to be thoroughly understood.\cite{ball2008water}
However, the understanding of liquid water stands by itself as a challenge in many aspects.
Unlike other liquids, water shows a lot of anomalies such as the density maximum at 4$^{\circ}$C,
the isobaric heat capacity minimum at 35$^{\circ}$C, and many more.\cite{gallo2016water,sun2014liquid}
Understanding the microscopic structures and dynamics of liquid water, in particular its hydrogen bond (HB) network,
is the key to understand these anomalies.\cite{palmer2014metastable,nilsson2011perspective,nilsson2010JESRP,13L-Swartz}
Recently, the high-resolution oxygen $K$-edge core-level spectroscopy, such as X-ray
absorption spectra (XAS) and X-ray Raman scattering,
has emerged as a powerful experimental technique
to probe the electronic structure and infer the molecular structure of water and ice, as well as aqueous solutions.
%to detect the structure of water and ice, as well as aqueous solutions.
\cite{Wernet995,chen2010PRLxas,kong2012PRB,tse2008PRL,head2006tetrahedral,smith2006probing,nordlund2009sensitivity,vinson2012theoretical,
fuchs2008isotope,giulia2006PRLxas,nilsson2010JESRP,nilsson2010x,fransson2016requirements,sellberg2014comparison,schreck2016isotope,nagasaka2017reliable}
Excited from the oxygen 1$s$ core level, the electron excitation probes the unoccupied electronic states,
which are anti-bonding along the covalent OH bonds and particularly sensitive to the HBs.
Therefore, the XAS technique serves as a local probe for the HB structures of liquid water and ice.
%The XAS technique explores the excited electronic state by exciting the electron from the oxygen $1s$ core level,
%serving as a very sensitive element-specific local probe for the HB structures of liquid water and ice.

The experimental XAS of water have three distinct features
as a function of increasing excitation energies:
a pre-edge starts from the absorption threshold at 533 to 536 eV with a peak centered at 535 eV,
a main-edge spanning from 537 to 539 eV and
a post-edge from 539 eV and beyond.\cite{tse2008PRL,nilsson2010x,sellberg2014comparison,schreck2016isotope,nagasaka2017reliable}
Experimentally, the pre-edge feature is present in both liquid water and ice
%but is more significant in the former.
but is more intense in the former.
The relative intensities between main-edge and post-edge
of liquid water and crystalline ice Ih are substantially different.
\cite{Wernet995,chen2010PRLxas,kong2012PRB,tse2008PRL,sellberg2014comparison,nilsson2010JESRP,pylkkanen2010role,sahle2013microscopic,schreck2016isotope,nagasaka2017reliable}
As one of the most qualitative differences, the intensity of main-edge is higher than that of the post-edge
in the XAS of liquid water,
while the opposite trend is true for the spectra in the ice.

The unambiguous assignments of the XAS features to the underlying HB structures are prerequisites to
the physical interpretation of the experimental spectra,
which can be achieved by first-principles methods including
both the modeling of
molecular structure of liquid water and the electron-hole excitation process.
%Given that the electronic ground state is generated by density functional theory (DFT),\cite{hohenberg64,kohn65}
%the {\it ab initio} molecular dynamics (AIMD)\cite{85L-CPMD} provides an ideal theoretical approach in 
%modeling liquid structure.
With the snapshots of represented molecular configurations from an equilibrated molecular dynamics trajectory,
the XAS can be computed with the knowledge of electronic structures
of the excited core-hole.
In this regard, various approximations\cite{hetenyi2004calculation,cavalleri2004x,giulia2006PRLxas}
for excited electronic states have been proposed within the framework of density functional theory (DFT).\cite{hohenberg64,kohn65}
%such as full core-hole (FCH), half core-hole, and excited electron and core-hole approximations.\cite{hetenyi2004calculation,cavalleri2004x,giulia2006PRLxas}
%The theoretical modeling of XAS in liquid water and ice was pioneered by
%Prendergast et al.\cite{giulia2006PRLxas}
In the seminal work of Prendergast et al.\cite{giulia2006PRLxas}, the proposed excited electron and core-hole approximation
yielded XAS in close agreement with experimental measurements.
More rigorously, the XAS of liquid water can be computed by solving the Bethe-Salpeter equation (BSE)
describing the electron-hole interaction.\cite{rehr2005final,vinson2011bethe,vinson2012theoretical}
The BSE approach involves the calculations of self-energy operator and quasiparticles,
which are in general computationally expensive for liquid water.
An approximate way of solving the BSE with less computational resources
was introduced by Chen et al.\cite{chen2010PRLxas}
based on a model electron screening function in the static Coulomb hole and screened exchange (COHSEX) approximation\cite{onida2002electronic},
which was evaluated using the maximally localized Wannier functions as basis,\cite{09B-Wu}
reducing the computational cost significantly.
In the above work, the molecular origins of the spectra features as well as 
the intensity difference between the main- and post-edge in water and ice has been validated\cite{chen2010PRLxas}.

Despite these recent developments, two uncertainties still remain
in the theoretical XAS of liquid water obtained by the approximate solutions of BSE.
The first uncertainty comes from the drawback in modeling the liquid water structure by
employing the generalized gradient approximation (GGA) in the framework of 
DFT.\cite{asthagiri2003free,grossman2004towards,schwegler2004towards,vandevondele2005influence,yoo2011communication,mogelhoj2011ab,wang2011density,yoo2009phase,zhang2011first,zhang2011structural,distasio2014individual}
It is known that GGA predicts an over-structured liquid water.\cite{yoo2009phase,zhang2011first,zhang2011structural,distasio2014individual}
Thus, a significantly elevated temperature is often
adopted to generate a softer HB structure closer to
the experimental measurement.\cite{zhang2011first,zhang2011structural,distasio2014individual}
The lack of physics in accurately describing the water structure
is due to the neglected van der Waals (vdW) interactions and
the spurious self-interaction error\cite{perdew1981self} in GGA functional.
Specifically, by including the vdW interactions, the water population
in the interstitial region between the first and second coordination shells of water molecules
is increased to better match experiment.\cite{distasio2014individual}
Furthermore, by mitigating the self-interaction error through the hybrid functional,
the directional HB strength between water molecules are weakened towards the experiment;
as a result, the protons are less easily to be donated to neighboring water molecules.\cite{distasio2014individual}
However, the effects of this improved water structure on the theoretical XAS have not been elucidated.
%Conventionally, a significantly elevated temperature is often adopted to generate a softer HB structure
%closer to the experimental measurement.
Secondly, in the series of computational work adopting the static COHSEX approximation for
the electron-hole excitation of liquid water, a homogeneous electronic screening model was first adopted\cite{chen2010PRLxas}
and then extended by using the Hybertsen-Louie ansatz\cite{88B-Mark} to
account for the inhomogeneous screening effects from the molecular environment.\cite{kong2012PRB}
However, some discrepancies still exist and it is not yet clear that to which extent the dynamic screening effect will affect the quasiparticle wavefunctions (QWs)
as well as the computed XAS.
For example, it was observed that the width of the theoretical XAS by static COHSEX is slightly narrower than
the experimental data.\cite{kong2012PRB}

In an effort to address the above issues,
we adopted a systematic way to study the XAS of liquid water at ambient conditions.
Specifically, we used more advanced {\it ab initio} modelings of molecular structures and electronic excitations.
We generated liquid water trajectories from 
the {\it ab initio} molecular dynamics (AIMD)\cite{85L-CPMD} simulations by employing a
hierarchy of XC functionals of PBE,\cite{perdew1996generalized}
PBE with the vdW interactions in the form of Tkatchenko and Scheffler (PBE+vdW),\cite{09vdW-TS} and
hybrid functional PBE0\cite{perdew1996rationale,adamo1999toward} with the vdW interactions (PBE0+vdW).
By utilizing the enhanced static COHSEX method to treat the excitations\cite{kang2010enhanced},
we find that the XAS computed from the snapshot generated by the PBE0+vdW functional
agree well with the experiment among the three XC functionals studied.
The vdW interactions soften the water structures by increasing the population of water molecules in the interstitial region,
while the hybrid functional mitigates the self-interaction error and weakens the HB strength towards experiment.\cite{distasio2014individual}
Both structural corrections and improved excitation theory are crucial in giving rise to an
overall improvement of the three edges of XAS.
In particular, the post-edge feature in the high-energy region of XAS is
in slightly better agreement with experiment by a
better screening modeling as the enhanced static COHSEX.
%considering the dynamics screening effects.
%approximate energy-dependent corrections.
In addition, we also compared a set of XAS computed from different excitation theories to the experiment
in order to show the importance of self-consistently diagonalized QWs
in capturing the qualitative features of XAS.

\section{Methods}

We performed AIMD simulations to generate liquid water trajectories
using a modified version of the Quantum ESPRESSO package.\cite{codeQE}
We simulated 128 water molecules in a periodic cubic cell with a cell length of 15.68~\AA{} using
the Car-Parrinello molecular dynamics (CPMD)\cite{85L-CPMD} within the canonical (NVT) ensemble.
We employed the norm-conserving pseudopotentials in the form of Troullier-Martins\cite{troullier1991efficient}
and set the kinetic energy cutoff of the electronic wavefunctions as 71 Ry.
We used a hierarchy of XC functionals, including PBE, PBE+vdW, and PBE0+vdW as mentioned.
The hybrid functional PBE0 with a mixing of 25\% exact exchange
was evaluated in a linear-scaling manner by taking advantages of maximally localized Wannier functions.\cite{09B-Wu}
The ionic temperatures were controlled through
the Nos\'{e}-Hoover chain thermostats with a chain length of 4 for each ion.\cite{84JCP-Nose,85A-Hoover,92JCP-Martyna}
All AIMD simulations were performed at 330 K,
where an increase of 30 K has been found to mimic the nuclear quantum effect (NQE) in structural quantities such as the oxygen-oxygen radial distribution function
in DFT based simulations of liquid water.\cite{08L-Morrone}
A time step of 4.0 a.u. and a fictitious electron mass of 300 a.u. were chosen.

We calculated the X-ray absorption cross section using the Fermi's golden rule:
\begin{align}\label{eq1}
\sigma(\omega)=4\pi^{2}\alpha_{0}\hbar\omega\sum_{f} | M_{if} |^{2} \delta(\omega_{if} - \omega),
\end{align}
where $\alpha_{0}$ is the fine structure constant and
$\hbar\omega$ is the absorbed photon energy matching the energy difference ${ \hbar\omega_{if} = E_{f} - E_{i} }$.
$E_{f}$ and $E_{i}$ are the eigenvalues of the final and initial states, respectively.
$M_{if}$ are the transition matrix elements between the initial state $\ket{\phi_{i}}$
and final state $\ket{\phi_{f}}$ that can be evaluated within the electric-dipole approximation as
$M_{if} \sim \bra{\phi_{i}} x \ket{\phi_{f}}$,
averaging over the three Cartesian directions.
We take the 1$s$ atomic core wavefunction from DFT calculations as the initial state $\ket{\phi_{i}}$.
For the final state $\ket{\phi_{f}}$, we apply a self-consistently diagonalization procedure
within the enhanced static COHSEX approach\cite{kang2010enhanced} and the details are as follows.

The final state $\ket{\phi_{f}}$ is obtained by utilizing the enhanced COHSEX approach,
which has been implemented within the framework of CPMD\cite{85L-CPMD} scheme.
Specifically, the COHSEX method has been implemented in the CPMD module within the Quantum ESPRESSO package.\cite{codeQE}
For each set of input wave functions $\ket{\phi_{f}}$, we fix
the ion positions and damp the wave functions of a system in a self-consistent way as explained below.
Note that the formulas we describe here are only suitable for the excitation theory we adopt
and are independent of the other CPMD simulations we performed with vdW and PBE0 functionals.

First, the Lagrangian from the CP approach is
\begin{align}
\mathcal{L}&=\frac{\mu}{2}\sum_{i}\langle\dot{\psi}_{i}|\dot{\psi}_{i}\rangle
+\frac{1}{2}\sum_{I}M_{I}\dot{\mathrm{\bf R}}_{I}^2-E_{tot}(\mathrm{\bf R},\{\psi\})\\
&+\lambda_{ij}(\langle\psi_{i}|\psi_{j}\rangle-\delta_{ij}), \nonumber
\end{align}
where $\nu$ is a fictitious mass of electrons and $\psi_{i}$ is the orbital of state $i$.
$M_{I}$ is the mass for atom $I$ that is located at $\mathrm{\bf R}_{I}$.
$E_{tot}(\mathrm{\bf R},\{\psi\})$ is the total energy calculated from first-principles methods.
The last part is the orthogonality constrain imposed on orbitals with the Lagrangian multiplier to be $\lambda_{ij}$.
Note that the initial ion positions are fixed and only wave functions are generated.
Therefore, we need to damp the wave functions towards the ground state, which is realized via
the equations of motions of electrons in plane-wave basis set
\begin{align}
\mu\ddot{\psi}_{i}=-{\it H}(\mathrm{\bf R},\{\psi\})\psi_i+\sum_{j}\lambda_{ij}\psi_j,
\end{align}
respectively. Here ${\it H(\mathrm{\bf R},\{\psi\})}$ is the Hamiltonian matrix of the system.
Furthermore, the Hamiltonian part can be evaluated in real space as
\begin{align}
{\it H}\psi(\mathbf{r})&=[-\frac{1}{2}\nabla^2(\mathbf{r})+V_{ext}(\mathbf{r, R})+V_H(\mathbf{r})]\psi(\mathbf{r})\\
&+\int d\mathbf{r}^{\prime}\Sigma(\mathbf{r},\mathbf{r^{\prime}},E)\psi(\mathbf{r^{\prime}}),
\end{align}
where $V_{ext}(\mathbf{r, R})$ is the external potential and $V_H(\mathbf{r})$ is the Hartree potential. In particular,
$\Sigma(\mathbf{r},\mathbf{r^{\prime}},E)$ is the self-energy operator that is non-local in real space and depends on the self-energy $E$.
In the static COHSEX approximation, the self-energy operator can be approximated as
\begin{align}
\Sigma_{\mathrm{COHSEX}}^{static}(\mathbf{r,r^{\prime}})=\Sigma_{\mathrm{COH}}^{static}(\mathbf{r,r^{\prime}})+
\Sigma_{\mathrm{SEX}}^{static}(\mathbf{r,r^{\prime}}).
\end{align}
The first part is
\begin{align}
\Sigma_{\mathrm{COH}}^{static}(\mathbf{r},\mathbf{r^{\prime}})&=\frac{1}{2}\delta(\mathbf{r-r^{\prime}})W_p(\mathbf{r, r^{\prime}}; E=0)\\
&=\frac{1}{2}\delta(\mathbf{r-r^{\prime}})(W-v),
\end{align}
where $W_p$ is the Coulomb hole, $W$ is the screened Coulomb interaction, and $v$ is the bare Coulomb interaction.
The Hybertsen-Louie ansatz\cite{88B-Mark} proposed that $W$ generally follows the local charge density and
has a form of
%can be written as
\begin{align}
W(\mathbf{r},\mathbf{r^{\prime}};E=0)=\frac{1}{2}(W[\mathbf{r}-\mathbf{r^{\prime}};\rho(\mathbf{r}^{\prime})]
+W[\mathbf{r^{\prime}}-\mathbf{r};\rho(\mathbf{r})]),
\end{align}
where $W$ can be written as
\begin{align}
W[\mathbf{r^{\prime}}-\mathbf{r};\rho(\mathbf{r})]=
\frac{1}{2\pi^3}\int\epsilon^{-1}[\mathbf{q};\rho(\mathbf{r})]
v(\mathbf{q})e^{i\mathbf{q}\cdot(\mathbf{r}-\mathbf{r^{\prime}})}d\mathbf{q}.
\end{align}
Here, we use the Bechstedt model\cite{92bechstedt} for the dielectric function
\begin{align}
\epsilon[\mathbf{q},\rho(\mathbf{r})]&=1+\Bigl[(\epsilon_{0}-1)^{-1}\\
&+\alpha(q/q_{\rm{TF}})^2+q^4/(\frac{4}{3}k_{F}^2q_{\rm{TF}}^2)\Bigr]^{-1}, \nonumber
\end{align}
where $q_{\rm{TF}}$ is the Thomas-Fermi wave vector, $k_{\rm{F}}$ is the Fermi wave vector,
$\epsilon_{0}$ is taken from experiment, and $\alpha$ is fixed
by matching the Bechstedt model to the $q^2$ dependence of the Penn model.\cite{penn1962wave}
Next, we can transform $W$ to
\begin{align}
W[\mathbf{r^{\prime}}-\mathbf{r};\rho(\mathbf{r})]&=\frac{v(\mathbf{r}-\mathbf{r^{\prime}})}{\epsilon_{0}}
-\frac{1}{a(x_1-x_2)|\mathbf{r}^{\prime}-\mathbf{r}|}\\
&\times\Big(\frac{e^{i\sqrt{x_1}|\mathbf{r}^{\prime}-\mathbf{r}|}}{x_1}
-\frac{e^{i\sqrt{x_2}|\mathbf{r}^{\prime}-\mathbf{r}|}}{x_2}\Big). \nonumber
\end{align}
Here $x_{1,2}=(-b\pm\sqrt{b^2-4ac})/2a$, $a=(\frac{4}{3}k_{F}^2q_{\rm{TF}}^2)^{-1}$,
$b=\alpha/q_{\rm{TF}}^{2}$, and $c=\epsilon_0/(\epsilon_0-1)$.
The second part is
\begin{align}
\Sigma_{\mathrm{SEX}}^{static}(\mathbf{r}^{\prime},\mathbf{r})=-\sum_{i}^{occ}\psi_{i}(\mathbf{r})
\psi_{i}^{\ast}(\mathbf{r^{\prime}})W(\mathbf{r},\mathbf{r^{\prime}};E=0).
\end{align}
Numerically it has been shown that most of the error by using the static COHSEX approximation, when compared to the GW,
comes from the short-wavelength part of the assumed adiabatic accumulation of the Coulomb hole $W_{p}$, namely, the COH part;
while the SEX term in the static COHSEX approximation yields relatively close values to the GW calculations.\cite{kang2010enhanced}
Therefore, the enhanced static COHSEX was proposed to introduce a universal function $f$ to approximately
include the dynamics screening in the original static model
into the COHSEX formula.
Specifically, the enhanced static COH term can be evaluated as
\begin{align}
\Sigma_{\mathrm{COH}}^{new}(\mathbf{r},\mathbf{r^{\prime}})&=\frac{\delta(\mathbf{r-r^{\prime}})}{2}
\int W_{p}(q;E=0)f(q/k_{f})e^{-i\mathbf{q}\cdot\mathbf{r}}d\mathbf{q},
\end{align}
where $q$ is a plane wave and $k_f$ is the Fermi vector.
The scaling functions is
\begin{align}
f(x)=\frac{1+a_1x+a_2x^2+a_3x^3+a_4x^4+a_5x^5+a_6x^6}{1+b_1x+b_2x^2+b_3x^3+b_4x^4+b_5x^5+b_6x^6},
\end{align}
where $a_1=1.9085$, $a_2=-0.542572$, $a_3=-2.45811$, $a_4=3.08067$, $a_5=-1.806$, $a_6=0.410031$,
$b_1=2.01317$, $b_2=-1.55088$, $b_3=1.58466$, $b_4=0.368325$, $b_5=-1.68927$, and $b_6=0.599225$.

In the above, we described the procedures to compute XAS based on an excited water molecule in a snapshot.
We then excited every water molecule in the 128-molecule supercell in order to sample 
the different local environment of disordered liquid water structure.
In our cases, we found converged XAS after randomly exciting 64 water molecules in the snapshot.  
%and we found converged XAS after exciting 64 water molecules in the snapshot.
We note that in a previous study, 10 individual and uncorrelated snapshots of 32 water molecules were chosen
and only small differences were observed between these snapshots.\cite{giulia2006PRLxas}
Therefore, we consider one snapshot of a large cell is sufficient to yield meaningful results and
take a representative snapshot from each AIMD trajectory 
reflecting the equilibrated structure of liquid water to compute XAS.

Due to the different local environments in liquid water of each excited oxygen atom,
we adopted the core-hole energy shift of each excitation by following
Ref.~\onlinecite{pehlke1993evidence},
which is a standard approach to compute the core level shifts. 
We used Gaussian broadening of 0.4 eV for all spectra,
which has been used in previous calculated XAS of liquid water.\cite{chen2010PRLxas,kong2012PRB}
The computed XAS were aligned to the onset of the pre-edge (535 eV) and then
normalized to the same area of experimental data ranging from 533 to 546 eV.

To analyze the real-space locations of QWs in terms of 
the three edges in XAS of liquid water, we also define the one dimensional 
density of QW as
\begin{align}
\rho_{i}(r)=\int\int |\psi_{i}(r,\theta,\phi)^2|d\theta d\phi,
\end{align}
along the radial direction and the origin is taken to be the position of the 
excited oxygen with a core-hole.
In the equation, $i$ represents the index of conduction band with excitation 
energy $\varepsilon_{i}$. In order to present the different localizations of the QWs, the 
index $i$ is chosen so that $\varepsilon_i\in\lbrack 534.5,535.5 \rbrack$eV for 
QWs representing pre-edge of XAS, 
$\varepsilon_i\in\lbrack 537,539 \rbrack$eV for QWs representing main-edge of XAS,
and $\varepsilon_i\in\lbrack 540,542 \rbrack$eV for QWs representing post-edge of XAS.

\begin{figure}
  \includegraphics[width=8.5cm]{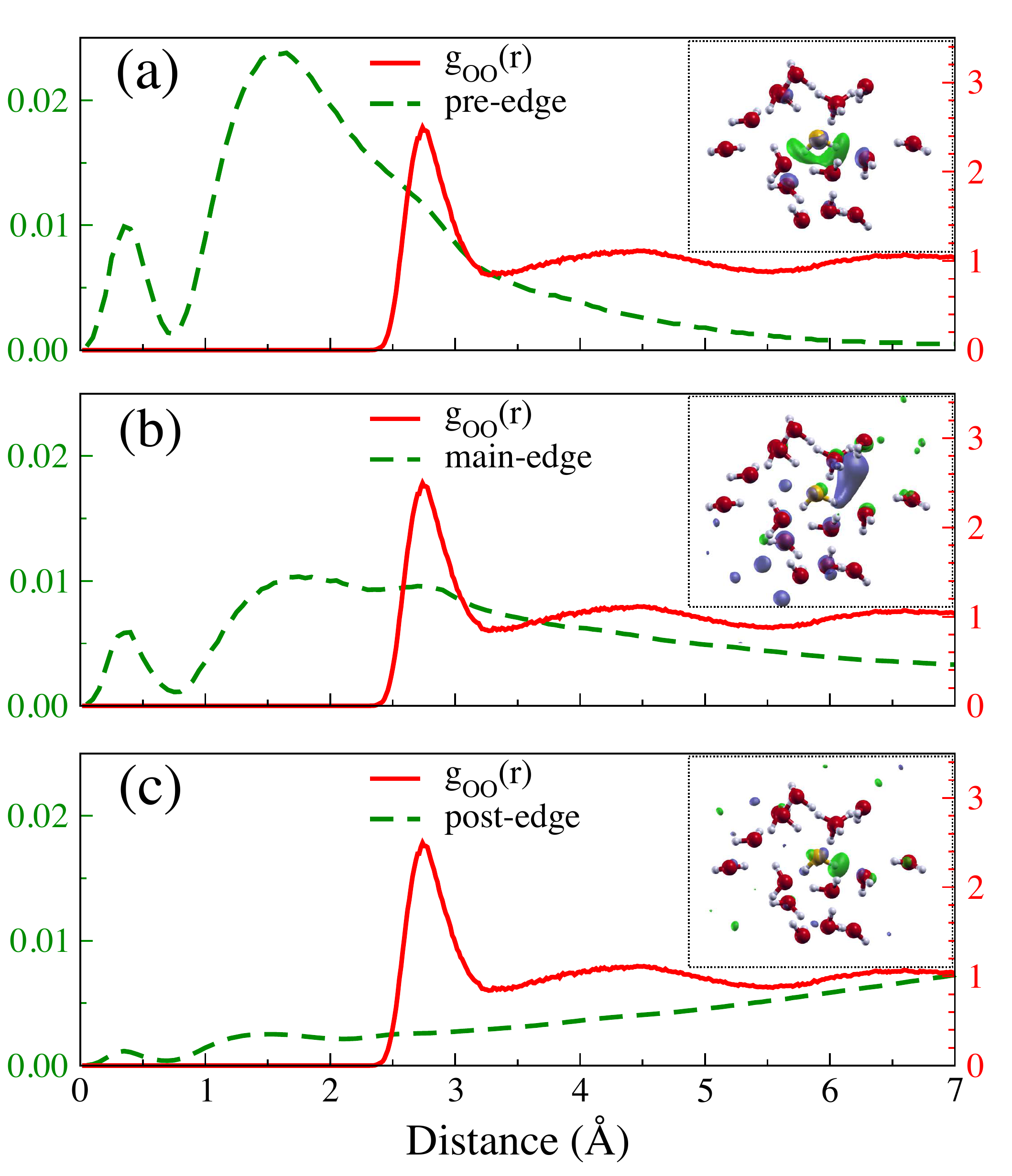}
  \caption{Density distributions of QW (green dashed line) of (a) pre-edge, (b) main-edge, and (c) post-edge
as a function of oxygen-oxygen distance
computed from the snapshot of the PBE0+vdW trajectory using the enhanced static COHSEX method.
%We averaged about top 20\% of the states with large transition amplitudes that contribute to the peaks of the three edges of XAS.
The $g_{\mathrm{OO}}(r)$ (red line) from the PBE0+vdW trajectory is shown for comparison in (a), (b), and (c).
The insets show the representative QWs of the three edges around the excited water molecule.
Water molecules reside within the second coordination shell of the excited oxygen are shown.
Red, white, and yellow spheres represent oxygen, hydrogen, and oxygen atom with a core hole, respectively.
The QWs with opposite signs are depicted with blue and green.
}
  \label{figwf}
\end{figure}

\section{Results and Discussion}

\subsection{HB structure probed by pre-, main- and post-edge of XAS}

The QWs can be strongly perturbed by the local liquid water structures.
The pre-edge, main-edge, and post-edge of XAS are found to
have distinguishable molecular signatures that
relate to different spatial regions of the HB network;\cite{Wernet995,chen2010PRLxas,nilsson2010x}
the pre-edge has the $4a_1$ character while
the main-edge and post-edge have the $b_2$ character, all of which
originate from the molecular excitations in the gas phase.\cite{Wernet995}
In order to quantitatively study the spatial regions in terms of different XAS edges,
we present the density distributions of QW as a function of oxygen-oxygen distance in Fig.~\ref{figwf}.
The oxygen-oxygen radial distribution function ($g_{\mathrm{OO}}(r)$)
and an excited oxygen with QW distributed within the HB network (insets) are also shown for comparison.
Overall, Figs.~\ref{figwf}(a), (b), and (c) show that the density distributions of QW
become more delocalized from the pre-edge, main-edge, to post-edge.

The density distribution of QW of pre-edge illustrated in
Fig.~\ref{figwf}(a) has the highest peak at 1.7~\AA~and is mostly localized within 2.75~\AA,
the later of which is the first peak position of $g_{\mathrm{OO}}(r)$.
The inset in Fig.~\ref{figwf}(a) shows that the QW of pre-edge
resembles the first excited state of water molecule in the gas phase with 4$a_1$ symmetry.
In this regard, our result is consistent with a previous assignment\cite{chen2010PRLxas}
of the pre-edge to a bound exciton state, where
the electron orbital was found to be mostly localized within the first coordination shell.
Therefore, the pre-edge features can be largely affected by the short-range structures,
such as broken HBs and covalent bond strength, around the excited oxygens.
With the weakened HBs described by hybrid DFT functional (PBE0) and vdW interactions, the short-range HB network,
therefore, the computed pre-edge of XAS is expected to be improved in both energies and intensities.

\begin{figure*}
  \includegraphics[width=18cm]{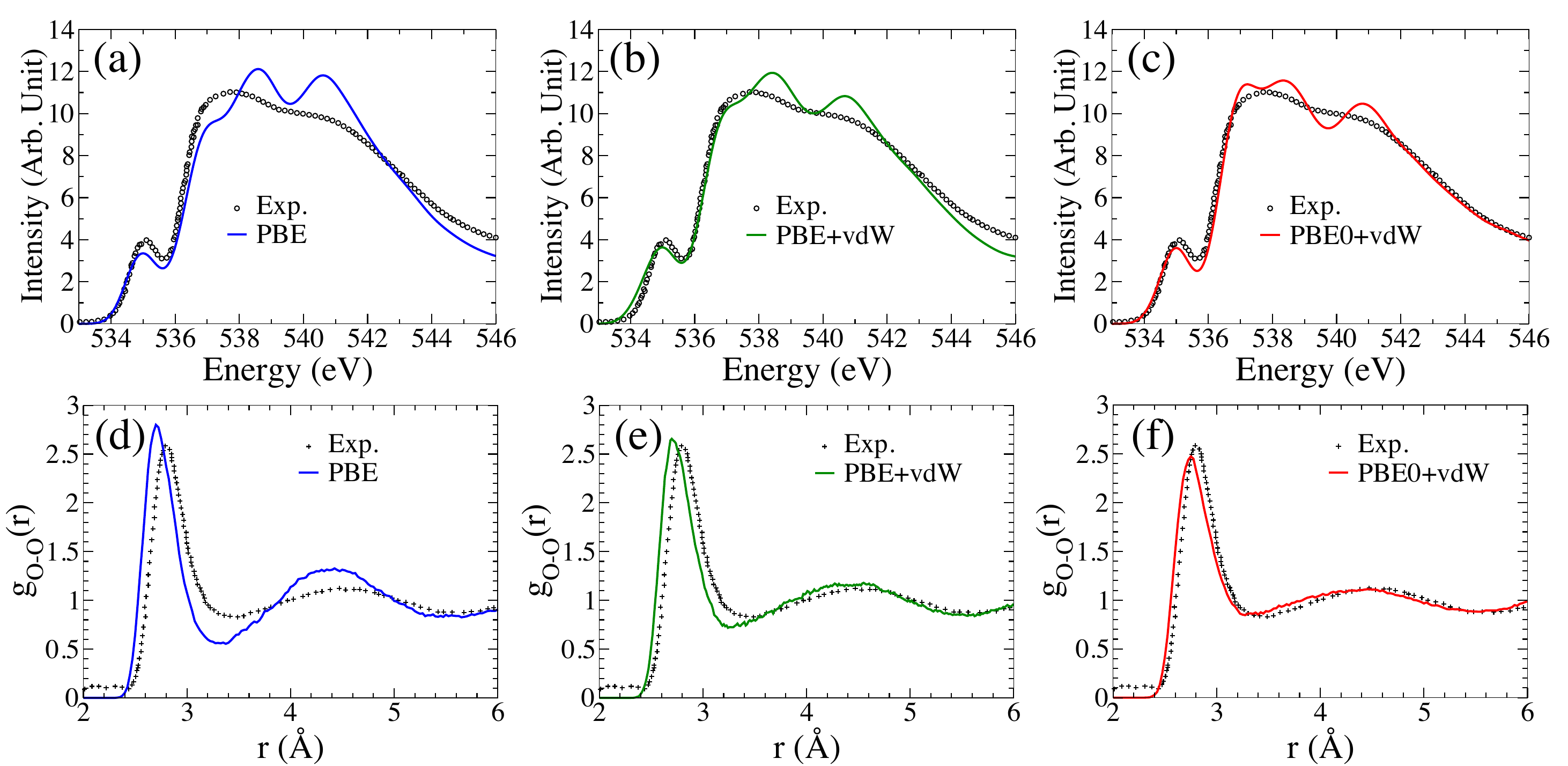}
  \caption{Computed XAS and $g_{\mathrm{OO}}(r)$ from three levels of 
  exchange-correlation functionals used in the AIMD simulations, namely,
PBE, PBE+vdW, and PBE0+vdW.
A representative snapshot consisting of 128 water molecules 
from each equilibrated AIMD trajectory was used for spectra calculation.
The enhanced static COHSEX method is adopted as the excitation theory.
The experimental (Exp.) data of XAS\cite{schreck2016isotope} 
and $g_{\mathrm{OO}}(r)$\cite{skinner2013benchmark} are also shown for comparison.
}
  \label{figAIMD}
\end{figure*}

As shown in Fig.~\ref{figwf}(b),
the density distribution of QW of the main-edge is more delocalized than that of the pre-edge.
The inset shows that the QW of main-edge can be found not only on the excited molecule itself but also on
its first and second shell neighbors.
In addition, a clear $b_2$ character can be identified for a typical QW of main-edge.
The above is consistent with the fact that main-edge was found to originate from the second excited state
of water molecule in the gas phase.
By comparing the localization of the main-edge density of QW
to that of the pre-edge, it can be seen that the former one is more localized between the first and second coordination shells of liquid water structure.
In this regard, it is expected that the main-edge feature of XAS will be sensitive to
the intermediate-range order of liquid
water structure, i.e. water molecules in the interstitial region.

In contrast to the density distributions of QW of the pre-edge and main-edge,
the QW of post-edge shown in Fig.~\ref{figwf}(c) is the most delocalized one but still preserves
$b_2$ character.
The strong delocalization can be clearly seen by the increased density of QW
as a function of the distance away from the excited water molecule.
Hence, the water molecules in long-range order
are critical in determining the post-edge features.
Using small simulation cell containing 32 water molecules fails to yield 
post-edge intensity of XAS in close agreement with experimental data.\cite{chen2010PRLxas}
The delocalization feature is consistent with the fact that the post-edge is a resonant exciton state.

\subsection{XAS calculated from PBE, PBE+vdW, and PBE0+vdW AIMD
trajectories}

Among the three levels of XC functionals investigated,
the XAS computed from the snapshot obtained with PBE show the least agreement with the experimental spectra in Fig. 2(a).
Four major discrepancies can be identified. First, the intensity of the computed pre-edge is
underestimated compared to experiment.
Second, the theoretical main-edge is centered around 538.5 eV, showing a large blue-shift (around 1 eV)
compared to the experimental value at 537.5 eV.
Third, a significantly over-estimated post-edge intensity leads to the fact
that the main-edge and post-edge have almost the same intensities,
which contradicts the experimental fact that the main-edge is more prominent than the post-edge in liquid water.
Fourth, the width of the computed XAS is slightly narrower than the experimental XAS, especially in the high-energy region.
Experimentally, the XAS of crystalline ice Ih with an intact HB network show a more prominent peak of post-edge than that of the main-edge.
Therefore, the discrepancies imply that the HB network of liquid water from the PBE AIMD trajectory
is over-structured.
Indeed, the $g_{\mathrm{OO}}(r)$ computed from the PBE AIMD trajectory significantly
deviates from experimental measurement\cite{skinner2013benchmark} as shown in Fig. 2(d).
Specifically, the first and second peaks of $g_{\mathrm{OO}}(r)$ from simulations are
significantly overestimated and the first minimum is largely underestimated.
Consistently, the average number of HBs per water molecule is found to be 3.76 in the PBE trajectory
according to the popular HB definition proposed by Luzar and Chandler,\cite{luzar1996effect}
which is the highest among all the functionals studied herein.
%Too many HBs and the overstructured RDF
All the above indicate that the HB network of liquid water is
over-structured from the PBE AIMD trajectory.
%We associate the over-structured HB network obtained from PBE causes
%the discrepancies mentioned above, i.e., a lower pre-edge and a very small difference between
%main-edge and post-edge intensities.
Not surprisingly, the theoretical XAS predicted by an over-structured HB network from the PBE AIMD trajectory
yield ice-like spectra with relatively more prominent post-edge feature.

\begin{figure}
  \includegraphics[width=8.5cm]{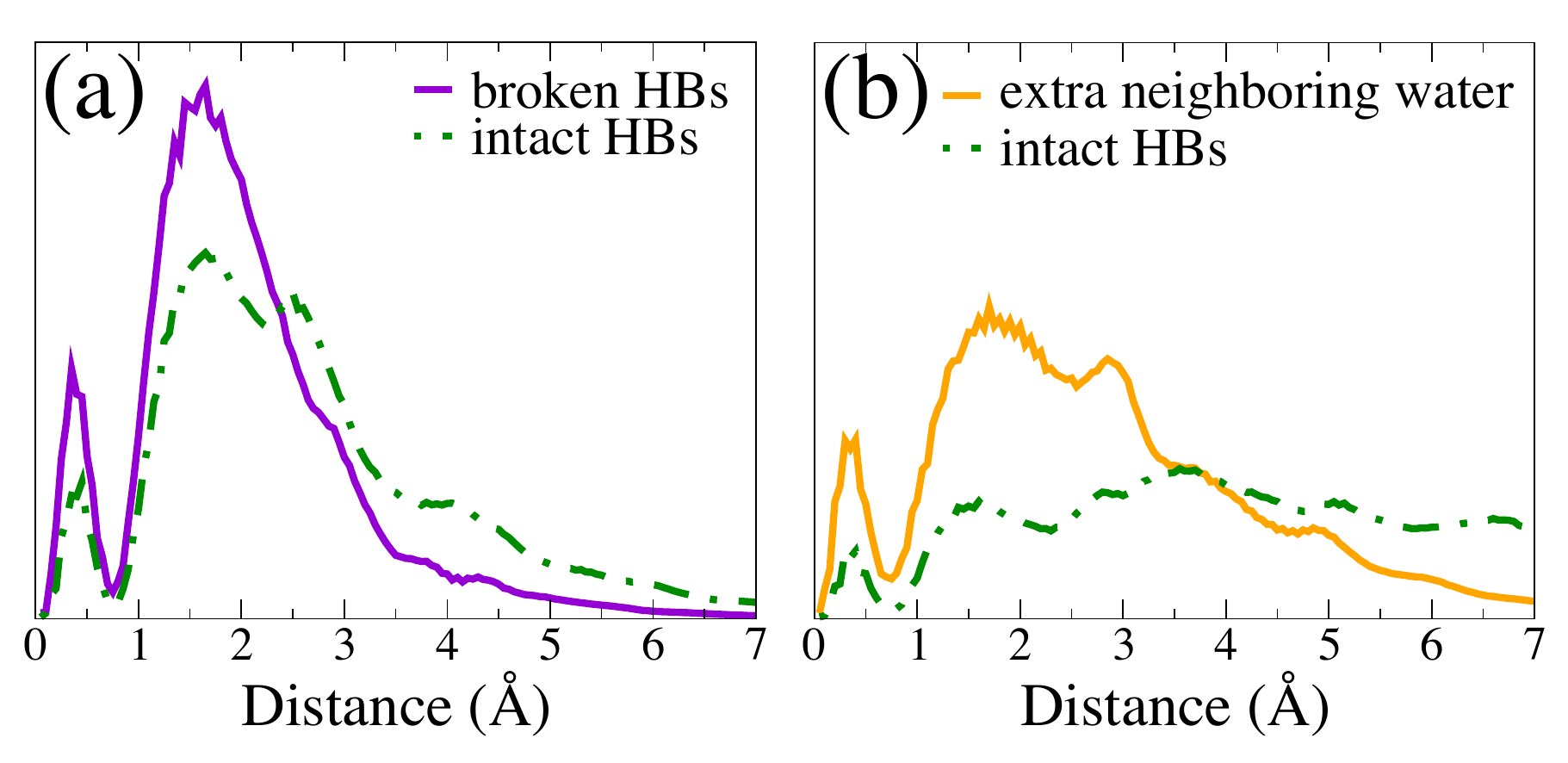}
  \caption{Density distributions of QW computed from different excited water molecules
 within a snapshot of 128 water molecules for the (a)
 pre-edge and (b) main-edge regions of XAS.
 The QWs are chosen based on different H-bonding environments and
 different electronic states.
 In panel (a), density distributions of QW are shown for
 two water molecules with an intact HBs (green dashed line) and broken HBs (purple solid line), respectively.
 Note the the intact HBs refer to two accepting and two donating HBs, and prominent peaks are observed for
 water molecules with broken HBs as shown in (a).
 In panel (b), density distributions of QW are shown for
 two water molecules with intact HBs.
 In particular, the presence of an extra non-bonded neighboring water molecules within
 the interstitial region (orange solid line) results in more prominent peaks as shown in (b).
}
  \label{fig3}
\end{figure}

As shown in Fig.~\ref{figAIMD}(b), the XAS computed from the PBE+vdW trajectory are largely improved towards the experimental spectra compared to that obtained from the PBE trajectory.
The improvement can be seen by a higher pre-edge intensity, a shift of main-edge towards lower energy, and a lower post-edge intensity.
The better agreement is attributed to the improved description of the HB network by including the vdW interactions in the AIMD simulation.
An explicit account of vdW forces strengthens the attractive interactions among water molecules,
which significantly increase the population of water molecules in the interstitial region.
The increased population of water molecules brings the $g_{\mathrm{OO}}(r)$ in closer agreement with experiment
within the first and second coordination shells.
Moreover, the increased population of water molecules in the interstitial region weakens the HBs among the water molecules in the first
coordination shell, resulting in a reduced first peak in the $g_{\mathrm{OO}}(r)$.
The average number of HBs per molecule is found to be 3.56, which is about 5\% smaller than that of PBE.
Hence, an excited oxygen atom experiences a more disordered environment by surrounding water molecules.
First of all, the pre-edge intensity is increased due to the breaking of more HBs.
In order to verify this, we selected two excited water molecules, one with broken HBs and the other one with four intact HBs,
and plotted their density distributions of QW in Fig.~\ref{fig3}(a).
The QW of pre-edge from the excited molecule with broken HBs is more localized with enhanced $p$ character
due to the more disordered short-range molecular environment.
Therefore, larger amplitudes of transition matrices $M_{ij}$ are obtained according to
the Fermi's golden rule as shown in Eq. (1).
As a result, the pre-edge intensity is increased with more broken HBs.
Secondly, the main-edge intensity is also enhanced by increased population of water molecules in the interstitial region
due to vdW interactions.
In order to further explain how the water molecules in the interstitial region affect the main-edge features,
we also selected two representative excited water molecules and calculated their density distributions of QW;
one of the excited water molecules has four intact HBs and
the other one has extra neighboring water molecules
within the interstitial region in addition to the four intact HBs.
% nearby interstitial water molecules while the other does not.
Fig.~\ref{fig3}(b) suggests that the density of the main-edge QW of the excited molecule with
extra neighboring water molecules within the interstitial region
% interstitial water
is more localized than the other by the disordered molecular environment in the intermediate range.
Similar to the discussion of the amplitudes of transition matrices and the QW localization
associated with the pre-edge feature,
we conclude that the increased population of water molecules within the interstitial region leads to larger amplitudes of transition matrices in the main-edge.
The above discussion explains the improved main-edge features, namely, the increased intensity and shift of the peak position (centering at 538.2 eV) towards experiments (centering at 537.5 eV).
The post-edge feature, whose QW is orthogonal to that of main-edge, is therefore predicted to have a lower spectra intensity.

The theoretical XAS are further improved towards the experimental measurements by using
the molecular configurations generated from the PBE0+vdW AIMD trajectory,
as shown in Fig.~\ref{figAIMD}(c).
By mixing a fraction of exact exchange in the hybrid functional,
PBE0 mitigates the self-interaction error
and lowers the tendency of hydrogens to be donated to neighboring water molecules.
Under the influence of exact exchange, the covalent bonds of water molecules are strengthened,
resulting in a shorter OH bond with the weakened directional HBs.
Consequently, the average number of HBs is reduced to 3.48 in the PBE0+vdW AIMD trajectory,
which is lower than the ones from the PBE and PBE+vdW AIMD trajectories.
As a result, the water structure is further softened with a larger fraction of broken HBs,
in consistence with the lower of the first peak of $g_{\mathrm{OO}}(r)$ as illustrated in Fig.~\ref{figAIMD}(f).
Moreover, the width of the first peak in $g_{\mathrm{OO}}(r)$ is broadened,
and the position is slightly increased towards the experimental direction,
in consistence with the weakened directional HB strength.
The revised HB network in liquid water essentially affects the computed XAS in the following three aspects.
First, the mitigated self-interaction error included in the hybrid functional
revises the short-range order of the HB network
in such a way that the excited water molecules experience a more disordered molecular environment comprising its first coordination shell,
which is sensitively probed by the XAS technique in the pre-edge region.
Based on the same argument in the previous paragraph,
the QW of pre-edge with broken HBs is more localized with enhanced $p$ character,
resulting in the more prominent pre-edge intensity with more broken HBs.
Therefore, the intensity of pre-edge from the PBE0+vdW trajectory is slightly increased towards the experimental spectra,
reflecting the softer liquid structure at short range.
Second, as a direct consequence of the weakened directional HB strength,
more non-bonded water molecules flow into the interstitial region as shown by the increased first minimum towards the experimental measure in Fig. 2(f).
As a result, the intermediate-range order of HB network is revised as well,
in which the excited water molecule is placed in a more disordered molecular environment beyond the first coordination shell.
Similar to the previous discussion of the effect of vdW interactions on main-edge, the main-edge features of XAS, which are sensitive to the intermediate-range order of the HB network, are modified with increased intensity ranging from 537 to 538 eV.
In addition, the peak position of main-edge is shifted to 537.9 eV, which is closer to the experimental peak at 537.5 eV.
The changes of the main-edge features provide better agreement with experimental spectra.
Third, the post-edge becomes more delocalized with a decreased intensity shifting towards higher energies.
The peak position of the post-edge is 540.8 eV, which is slightly smaller than the experimental value of 541.0 eV.
The more prominent feature of the main-edge than that of the post-edge is best captured from the PBE0+vdW trajectory.
In particular, the width of XAS is broadened in the high-energy region (544 to 546 eV), matching better with experiments.

\begin{figure}
  \includegraphics[width=8cm]{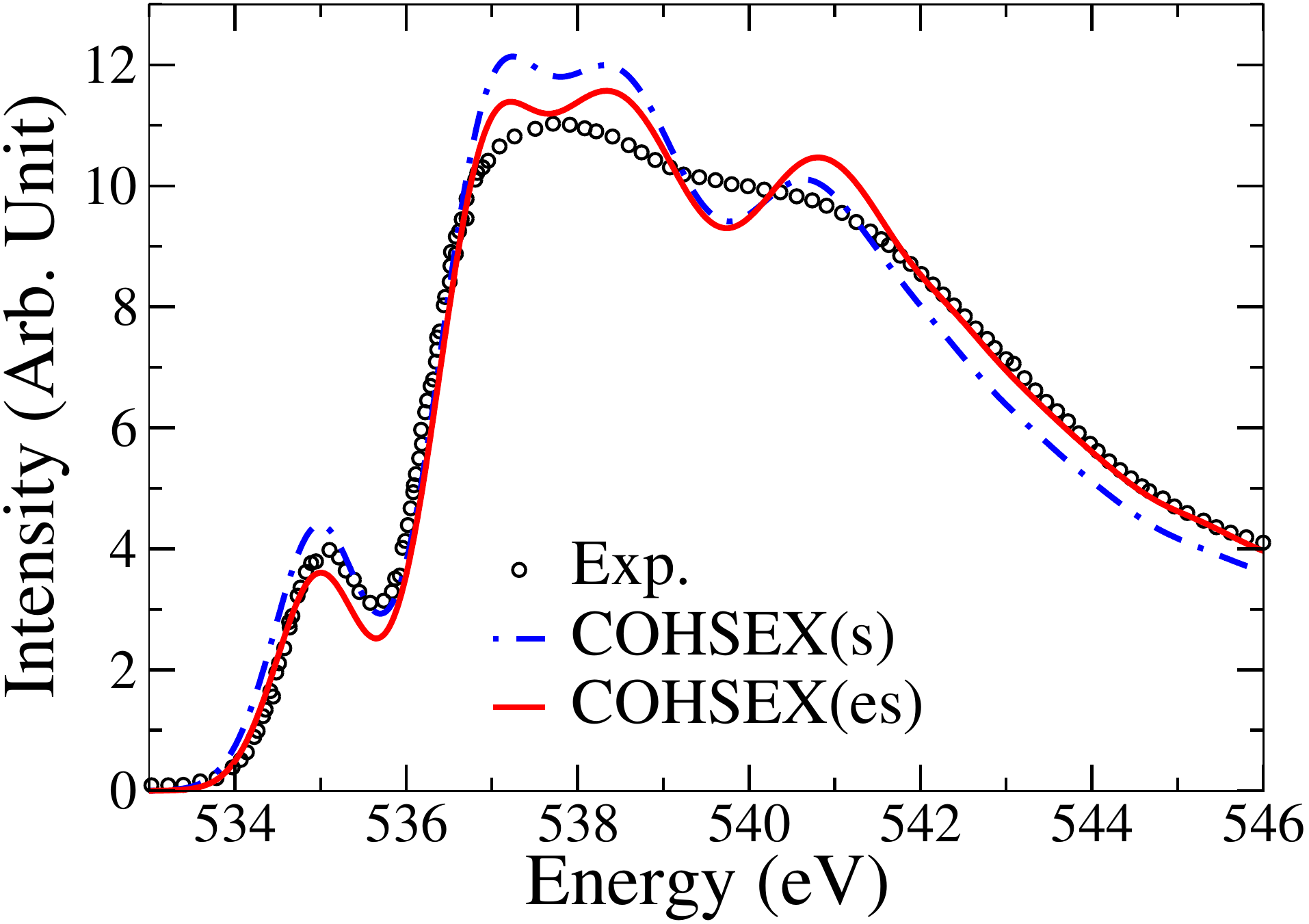}
  \caption{Comparison of XAS computed between static COHSEX (COHSEX(s)),
enhanced static COHSEX (COHSEX(es)), and the experimental data.\cite{schreck2016isotope}
A representative snapshot from the PBE0+vdW AIMD trajectory
containing 128 water molecules was used for all spectra calculations.
Self-consistent calculations were carried out to update the input wave functions.}
  \label{figenhance}
\end{figure}

\subsection{Enhanced Static COHSEX and Self-Consistently Diagonalized QWs}
Besides the accurate description of water structures by including
the vdW interactions and hybrid functional (PBE0),
a proper treatment of excitations is also critical in obtaining accurate XAS.
The approximation of the electron self-energy has been widely used in the electron excitation problems,
but is computationally very expensive in simulating XAS of liquid water.
In order to reduce the computational cost,
excitation theories based on the COHSEX method with electron screening models were used.\cite{chen2010PRLxas,kong2012PRB}
Compared to the GW methods, the static COHSEX method suffers from the wavelength dependent error from the COH part,
which is more significant at short wavelength but negligible at long wavelength.\cite{kang2010enhanced}
As a result, the high-energy region of XAS is affected but can be improved
by using the enhanced static COHSEX method.\cite{kang2010enhanced}

Fig.~\ref{figenhance} shows the XAS of liquid water computed by using both static COHSEX
and enhanced static COHSEX methods, and the water structure is chosen from the PBE0+vdW
trajectory. By using the enhanced static COHSEX method,
the XAS of liquid water are improved towards experiment when compared to
the spectra calculated from the static COHSEX.
For example, the intensity of the post-edge region ($>$ 541 eV)
increases while intensities of the pre-edge and main-edge decrease, as obtained from the enhanced static COHSEX.
This is due to the broadening effect introduced by considering the dynamic screening
in the enhanced static COHSEX approach.
The above broadening effect is more significant on the short-wavelength part,
as can be seen by the corrections on the XAS in relatively high excitation energy region.
By normalizing the XAS of both experiment and theory to the same area,
the increase of the high-energy spectra also leads to the decrease
of the pre-edge and main-edge intensities.

\begin{figure}
  \includegraphics[width=8cm]{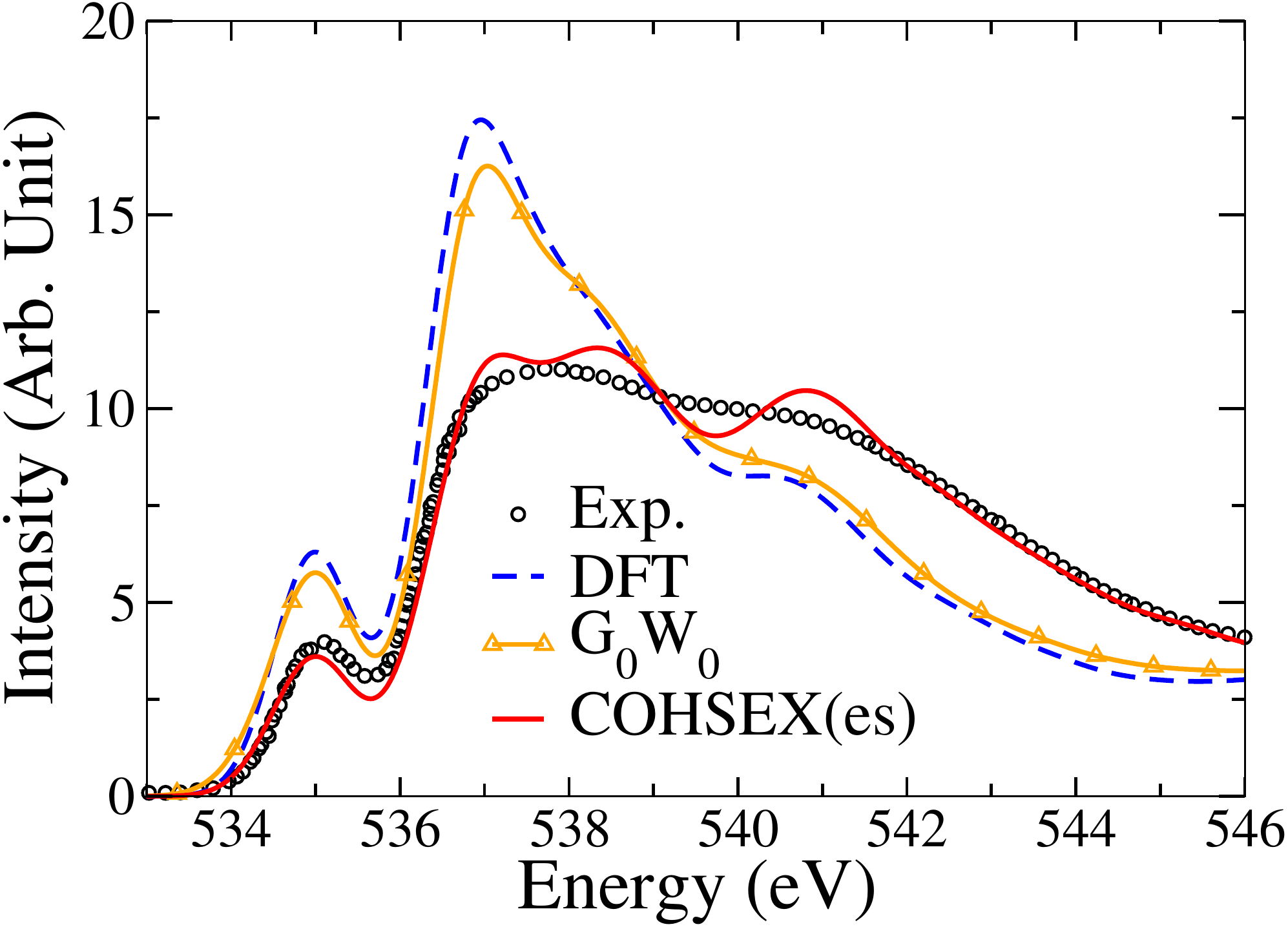}
  \caption{Comparison of XAS computed from three different excitation schemes
and the experiment.\cite{schreck2016isotope}
A representative snapshot from the PBE0+vdW AIMD trajectory
containing 128 water molecules was used for all spectra calculations.
The first excitation scheme by GGA-level DFT (blue dashed) with the FCH approximation is a one-step
calculation without changing the input wave functions.
Second, the G$_0$W$_0$ (orange) scheme was also performed with one-step diagonalizaiton.
Third, the enhanced static COHSEX (COHSEX(es)) (red solid line) was adopted
and a self-consistent calculation was carried out to update the input wave functions.
}
  \label{figscfGW}
\end{figure}

In order to emphasize the importance of self-consistently diagonalized QWs,
we further compare XAS as obtained from three different excitation treatments.
First, the full core-hole (FCH) approximation,
where both the energy levels and wavefunctions are from unoccupied Kohn-Sham eigenstates.
Second, a perturbative treatment based on the G$_0$W$_0$ approximation, in which the quasiparticle energies
are from the excitation theory but the QWs are still approximated by the Kohn-Sham eigen functions.
Third, the current enhanced static COHSEX approach in which
both the excitation energies and the QWs are generated by the self-consistently diagonalized self-energy operator.
%Due to the expensive computational cost,
%we computed all XAS based on a representative snapshot from the PBE0+vdW trajectory generated from
%a smaller cubic cell containing 32 water molecules.
The resulting spectra yielded from the above three different approaches are reported in Fig.~\ref{figscfGW}.
The G$_0$W$_0$ approximation has been widely used in calculating the
band structures of water,\cite{pham2014probing} aqueous solutions,\cite{opalka2014ionization} and many other
organic systems.\cite{blase2011first,marom2012benchmark,droghetti2014electronic,korzdorfer2012strategy,knight2016accurate,van2015gw}
However, calculations of the XAS significantly depend on the QWs as they are explicitly
involved in the evaluation of the transition matrix elements $M_{ij}$ in Eq. (1).
Not surprisingly, the use of the wavefunctions from the FCH approximation in DFT and G$_0$W$_0$ approach
leads to very similar predicted XAS.
As a result, XAS from both FCH and G$_0$W$_0$ approach yield a narrowed spectra width with spectra features significantly deviating from experiments as shown in Fig.~\ref{figscfGW}.
In sharp contrast, XAS from
self-consistently diagonalized self-energy operator within the enhanced static COHSEX approach are in
much better agreement with the experimental spectra.

\section{Summary}

We report a systematic modeling of the liquid water XAS
by advanced {\it ab initio} approaches.
We find that the computed XAS agree better with the experiment spectra
based on  the liquid structure generated from AIMD by
 including the vdW interactions and hybrid functional (PBE0).
The predicted XAS are further improved by the enhanced static COHSEX method
that includes the approximate dynamic screening.
%that adopts an approximate form of the energy dependency in the self-energy operator.
Specifically, the features of the three edges (pre-edge, main-edge, and post-edge) of liquid water are improved.
First, the short-range order in liquid water is improved by inclusion of the vdW interactions and PBE0 functional.
The vdW interactions increase the population of water molecules within the interstitial region
and therefore weaken the short-range HBs.
Under the influence of exact exchange by PBE0,
the covalent OH bonds are shorter, which further weakens the directional HB strength.
As a result, the pre-edge intensity is increased by the presence of more broken HBs in liquid water.
An accurate description of the intermediate-range order of water,
especially the water molecules in the interstitial region,
is crucial to yield the main-edge intensity and peak position close to experiment.
The inclusion of both vdW interactions and hybrid functional increases the population of
water molecules in the interstitial region, resulting in main-edge features that agree well with the experimental spectra.
Lastly, the overall spectra are improved by considering the dynamic screening effects
in the enhanced static COHSEX.
In conclusion, the PBE0+vdW AIMD trajectory with the enhanced static COHSEX yields better theoretical XAS.
QWs
from the diagonalization of the GW self-energy instead of the DFT wavefunctions
are crucial in obtaining spectra that quantitatively agree with experiments.

\section{acknowledgement}
The authors thank Roberto Car and Giulia Galli for helpful discussions.
This work was mainly supported by National Science Foundation (NSF),
DMR under Award DMR-1552287 (design of the project).
M.C. and M.L.K were supported by U.S. Department of Energy Scidac under Grant No. DE-SC0008726 (hybrid functional and van der Waals AIMD algorithms).
This research used computational resources of the National Energy Research Scientific Computing Center (NERSC),
a DOE Office of Science User Facility supported by the Office of Science of the U.S. Department of Energy under Contract No. DE-AC02-05CH11231. To whom correspondence should be addressed: xifanwu@temple.edu.

%\bibliography{mybibtex}

\end{document}